\documentclass[11 pt,twoside,aps,prd,amsmath,amssymb,
tightenlines,showpacs,showkeys]{revtex4-1}

\usepackage{amssymb}
\usepackage{amsmath}
\usepackage[dvips]{graphicx}
\usepackage{graphicx,epsfig}
\usepackage{bm}
\def\myfigure #1#2#3#4
{\begin{figure}[ht]\begin{center}
\includegraphics[width=#2 \textwidth]{#1.eps}
\parbox[t]{#4\textwidth}{\caption{#3}\label{#1}}
\end{center}\end{figure}}

\def \myfigures #1#2#3#4#5#6#7#8
{\begin{figure}[ht]
    \begin{center}
        \includegraphics[width=#2 \textwidth]{#1.eps}
        \hfill
        \includegraphics[width=#6 \textwidth]{#5.eps}
        \parbox[t]{#4\textwidth}{\caption {#3}\label{#1}}
        \hfill
        \parbox[t]{#8\textwidth}{\caption {#7}\label{#5}}
    \end{center}
\end{figure} }

\begin{document}
\baselineskip -24pt
\title{An exact Bianchi V cosmological model in Scale Covariant theory of gravitation: A variable deceleration parameter study}

\author{Mohd.Zeyauddin$^{1}$}, \author{C. V. Rao$^{2}$}
\affiliation{$^{1,2}$Department of General Studies (Mathematics), Jubail Industrial College, Jubail Industrial City, Kingdom of Saudi Arabia}
\email{mdzeyauddin@gmail.com}
\email{drcvrao1968@gmail.com}
\vspace{0.5cm}
\begin{abstract}
\noindent \textbf{Abstract.} A spatially homogeneous and anisotropic Bianchi type V cosmological model of the universe for perfect fluid within the framework of Scale covariant theory of gravitation proposed by Canuto et al., is studied in view of a variable deceleration parameter which yields the average scale factor $a(t)=sinh^{1/n} (\beta t)$, where $n$ and $\beta$ are positive constants. The solution represents a singular model of the universe. All physical and geometrical properties of the model are thoroughly studied. The time dependent deceleration parameter supports the recent observation. The model represents an accelerating phase for $0<n\leq1$ and for $n>1$, there is a phase transition from early deceleration to a present accelerating phase.
\end{abstract}

\maketitle

\pagebreak
\pagestyle{headings}
\section{Introduction}

\noindent The scalar tensor theories are the generalizations of Einstein's general theory of relativity in which the metric is generated by a scalar gravitational field together with non-gravitational field. The scalar gravitational field itself is generated by the non-gravitational fields via a wave equation in curved space-time. Nowadays scalar tensor theories are more in use as the Einstein's theory of general relativity doesn't seem to resolve some of the important problems in cosmology such as dark matter or the missing matter[1-13]. Scalar tensor theories provide a convenient set of representations for the observational limits on possible deviations from general relativity. Canuto et al.[14, 15] have formulated Scale covariant scalar tensor theory of gravitation by associating the mathematical operation of scale transformation with the physics of using different dynamical systems to measure space time distances. In this theory the generalized Einstein's field equations are invariant under scale transformations. This theory also provide a natural interpretation of the variation of the gravitational constant $G$ [16, 17]. Also in this theory, the Einstein's field equations are valid in gravitational units and the other physical quantities are measured in atomic units. The line element $d\bar{s}^{2}=\bar{g_{ij}}dx^{i}dx^{j}$ in Einstein units corresponds to $ds=\phi^{-1}(x)d\bar{s}$ in any other units (in atomic units). The metric tensor in the two systems of units are related by a conformal transformation $\bar{g_{ij}}=\phi^{2}g_{ij}$, where the metric $\bar{g_{ij}}$ giving macroscopic metric properties and $g_{ij}$ giving microscopic metric properties. Here we consider the guage function $\phi$ as a function of time.\\

\vspace{0.5cm}

\noindent Several workers in the field of general relativity and cosmology have studied the Scale-Covariant theory in different Bianchi space-times. Shri Ram et al. [18] have studied a spatially homogeneous Bianchi type V cosmological model in Scale-Covariant theory of gravitation. Zeyauddin et al. [19] have studied Scale covariant theory for Bianchi type VI space-time and obtained some exact solutions. Reddy et al. [20] have developed
a cosmological model with negative constant deceleration parameter in Scale-Covariant theory of gravitation. Reddy et al.[21] have presented the exact Bianchi type II, VIII and IX cosmological models in Scale-Covariant theory of gravitation. Beesham [22] has obtained a solution for Bianchi type I cosmological model in the Scale-Covariant theory. Higher dimensional string cosmologies in Scale-Covariant theory of gravitation have been investigated by Venkateswarlu and Kumar [23].

\vspace{0.2cm}

\section{Basic Equations}
\noindent The generalized field equations in Scale covariant theory of gravitation are given as

\begin{equation}
R_{ij}-\frac{1}{2}g _{ij}R+f_{ij}(\phi) =-8\pi G T_{ij}+\Lambda (\phi)g_{ij},
\end{equation}
where

\begin{equation}
\phi^{2}f_{ij}= 2\phi \phi_{i;j}-4\phi_{i}\phi_{j}-g_{ij}(\phi\phi^{\lambda}_{;\lambda}-\phi^{\lambda}\phi_{\lambda}),
\end{equation}
for any scalar $\phi, \phi_{i}=\phi_{,i}$. Here comma denotes ordinary differentiation whereas a semi-colon denotes a covariant differentiation. The line element for the Bianchi V space-time can be written as

\begin{equation}
ds^2=dt^2-A^{2}dx^2-e^{2m x}[B^{2}dy^2+C^{2}dz^2]
\end{equation}
where the scale factors $A$, $B$, $C$ are functions of time and $m$ is an arbitrary constant. Here we take the source of gravitational field as a perfect fluid. The energy momentum tensor  for a perfect fluid, is given by

\begin{equation}
T_{ij} = \left(\rho +p \right)u_{i}u_{j} - p g_{ij},
\end{equation}
where $\rho$ is the energy-density, $p$ the pressure and $u^{i}$ is four velocity vector of the fluid following $u^{i}u_{j}=1$.\\
\vspace{0.2cm}

\noindent Some basic physical parameters for the line element equation (3) are given by,
\begin{equation}
V=ABC,
\end{equation}

\begin{equation}
a=(ABC)^{1/3},
\end{equation}

\begin{equation}
\theta = u ^{\mu}_{;\mu}=\frac{\dot{A}}{A}+\frac{\dot{B}}{B}+\frac{\dot{C}}{C},
\end{equation}

\begin{equation}
\sigma ^{2} = \frac {1}{2}\sigma _{\mu\nu}\sigma^{\mu\nu}= \frac {1}{2}\left[\left(\frac{\dot{A}}{A}\right)^{2}+  \left(\frac{\dot{B}}{B}\right)^{2}+\left(\frac{\dot{C}}{C}\right)^{2}\right]-\frac{{\theta}^{2}}{6},
\end{equation}

\begin{equation}
H= \frac{1}{3}\left(H_{1}+H_{2}+H_{3}\right),
\end{equation}
where $V$, $a$, $\theta$, $\sigma^{2}$, and  $H$ are volume scalar, scale factor, expansion scalar, shear scalar, and  Hubble parameter respectively. Here $H_{1}=\frac{\dot{A}}{A}$, $H_{2}=\frac{\dot{B}}{B}$ and $H_{3}=\frac{\dot{C}}{C}$ are directional Hubble parameters in the directions of $x$, $y$ and $z$ respectively. A dot denotes the derivative with respect to time.  A very important relation in terms of the parameters $H$, $V$ and $a$ can be obtained as

\begin{equation}
H=\frac{1}{3}\frac{\dot{V}}{V}=\frac{\dot{a}}{a}.
\end{equation}
From equations (1)-(4), the field equations can be translated in the following set of non-linear differential equations,

\begin{eqnarray}
\frac{\ddot{B}}{B}+\frac{\ddot{C}}{C}+\frac{\dot{B}}{B}\frac{\dot{C}}{C}-\frac{m^2}{A^{2}}-2\frac{\dot{A}}{A}\frac{\dot{\phi}}{\phi}
+\frac{\dot{\phi}}{\phi}\frac{\dot{V}}{V}+\frac{\ddot{\phi}}{\phi}-\frac{\dot{\phi}^{2}}{\phi^{2}}=-8\pi G p,
\end{eqnarray}

\begin{eqnarray}
\frac{\ddot{A}}{A}+\frac{\ddot{C}}{C}+\frac{\dot{A}}{A}\frac{\dot{C}}{C}-\frac{m^2}{A^{2}}-2\frac{\dot{B}}{B}\frac{\dot{\phi}}{\phi}
+\frac{\dot{\phi}}{\phi}\frac{\dot{V}}{V}+\frac{\ddot{\phi}}{\phi}-\frac{\dot{\phi}^{2}}{\phi^{2}}=-8\pi G p,
\end{eqnarray}

\begin{eqnarray}
\frac{\ddot{A}}{A}+\frac{\ddot{B}}{B}+\frac{\dot{A}}{A}\frac{\dot{B}}{B}-\frac{m^{2}}{A^{2}}-2\frac{\dot{C}}{C}\frac{\dot{\phi}}{\phi}
+\frac{\dot{\phi}}{\phi}\frac{\dot{V}}{V}+\frac{\ddot{\phi}}{\phi}-\frac{\dot{\phi}^{2}}{\phi^{2}}=-8\pi G p,
\end{eqnarray}

\begin{eqnarray}
\frac{\dot{A}}{A}\frac{\dot{B}}{B}+\frac{\dot{A}}{A}\frac{\dot{C}}{C}+\frac{\dot{B}}{B}\frac{\dot{C}}{C}-\frac{3m^{2}}{A^{2}}
+\frac{\dot{\phi}}{\phi}\frac{\dot{V}}{V}-\frac{\ddot{\phi}}{\phi}+3\frac{\dot{\phi}^{2}}{\phi^{2}}=8\pi G \rho,
\end{eqnarray}

\begin{eqnarray}
2\frac{\dot{A}}{A}-\frac{\dot{B}}{B}-\frac{\dot{C}}{C}=0.
\end{eqnarray}
The continuity equation $T^{i}_{j;i}u^{j}=0$ reads,

\begin{eqnarray}
\dot{\rho}+(\rho+p)\frac{\dot{V}}{V}+\rho\left(\frac{\dot{\phi}}{\phi}+\frac{\dot{G}}{G}\right)+3p\frac{\dot{\phi}}{\phi}=0.
\end{eqnarray}
From equations (11)-(14), we obtain the energy density and pressure as follows:

\begin{eqnarray}
8 \pi G\rho= 3 H^2-\sigma^2 -\frac{3m^2}{A^2}-\frac{\ddot{\phi}}{\phi}+3 \left(\frac{\dot \phi}{\phi}\right)^2+3H\frac{\dot \phi}{\phi},
\end{eqnarray}

\begin{eqnarray}
8 \pi G p= H^2(2q-1)-\sigma^2 +\frac{m^2}{A^2}-\frac{\ddot{\phi}}{\phi}+ \left(\frac{\dot {\phi}}{\phi}\right)^2 - H\frac{\dot \phi}{\phi}.
\end{eqnarray}
From equation (15), we have

\begin{eqnarray}
A^2=BC.
\end{eqnarray}
Following the approach of Saha and Rikhvitsky [24], Zeyauddin and Saha [25] , Shri Ram et al.[26- 28], and Singh et al.[29],  we solve the equations from (11) to (14) and (19) to get the quadrature solution for the metric functions $A$, $B$ and $C$ as
\begin{equation}
A(t)=a,
\end{equation}
\begin{equation}
B(t)=B_0 a  \exp\left(M \int{\frac{dt}{a^3 \phi^2}}\right),
\end{equation}
\begin{equation}
C(t)=C_0 a  \exp\left(-M\int{\frac{dt}{a^3 \phi^2}}\right),
\end{equation}
where $B_0$, $C_0$ and $M$ are arbitrary constants. In this paper, we obtain exact solution to the field equations of Scale-Covariant theory for Bianchi type V space-time metric, using the concept of variable deceleration parameter. For that we follow the assumptions made by Pradhan[30], N. Ahmad et al.[31], Pradhan and Otarod[32], Akarsu and Dereli[33], Pradhan et al.[34], Chawla et al.[35], Chawla and Mishra[36], Mishra et al.[37], Pradhan [38], Pradhan et al. [39], Pradhan et al. [40], Amirhashchi et al.[41].  The deceleration parameter can be time dependent and for that the average scale factor is given by the following relationship as
\begin{equation}
a(t)=\left[\sinh{(\beta t)}\right]^{\frac{1}{n}}.
\end{equation}

\myfigures{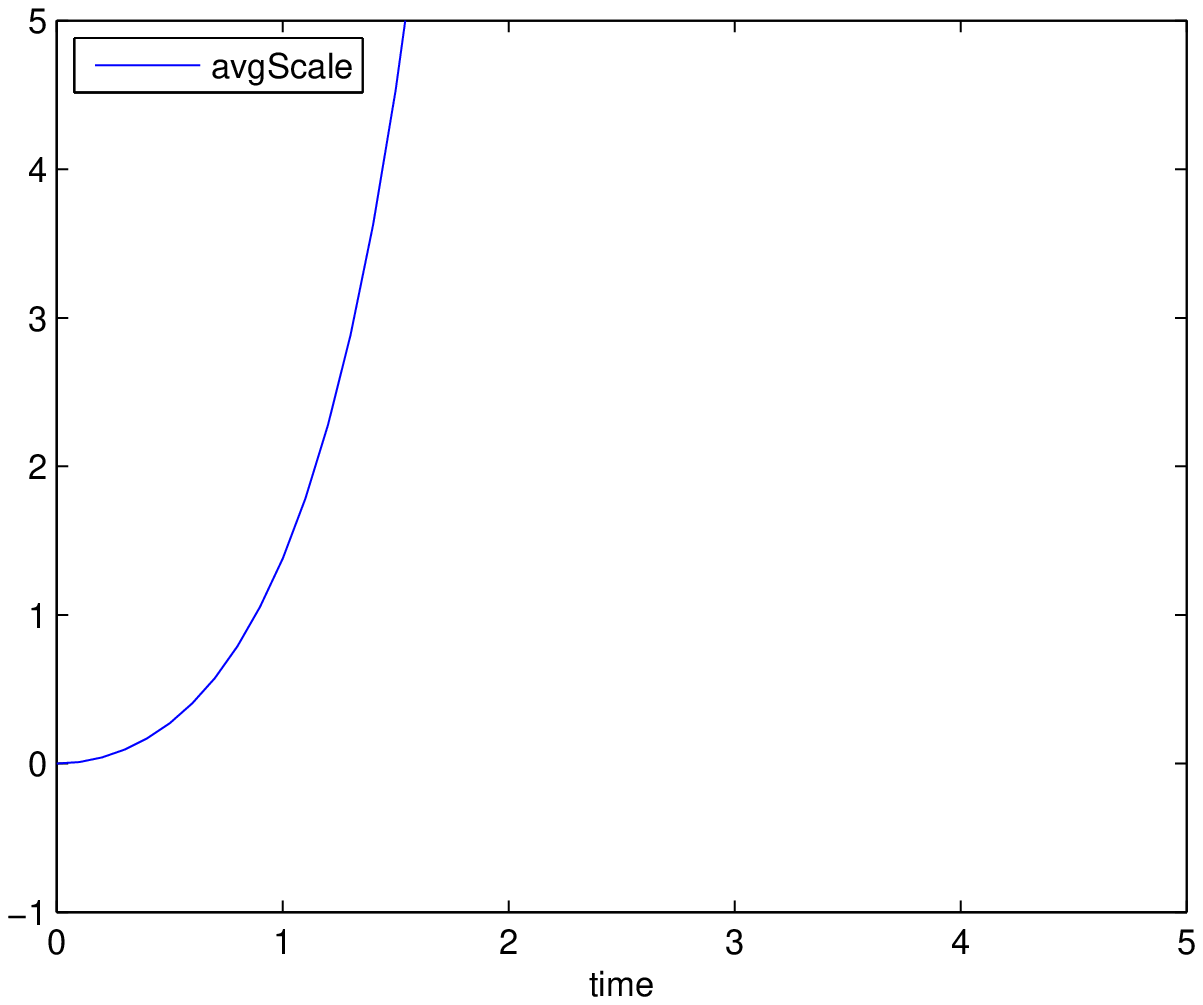}{0.45}{Variation of average scale factor $a$  for the
parameter $n=0.5$ with time.} {0.45}{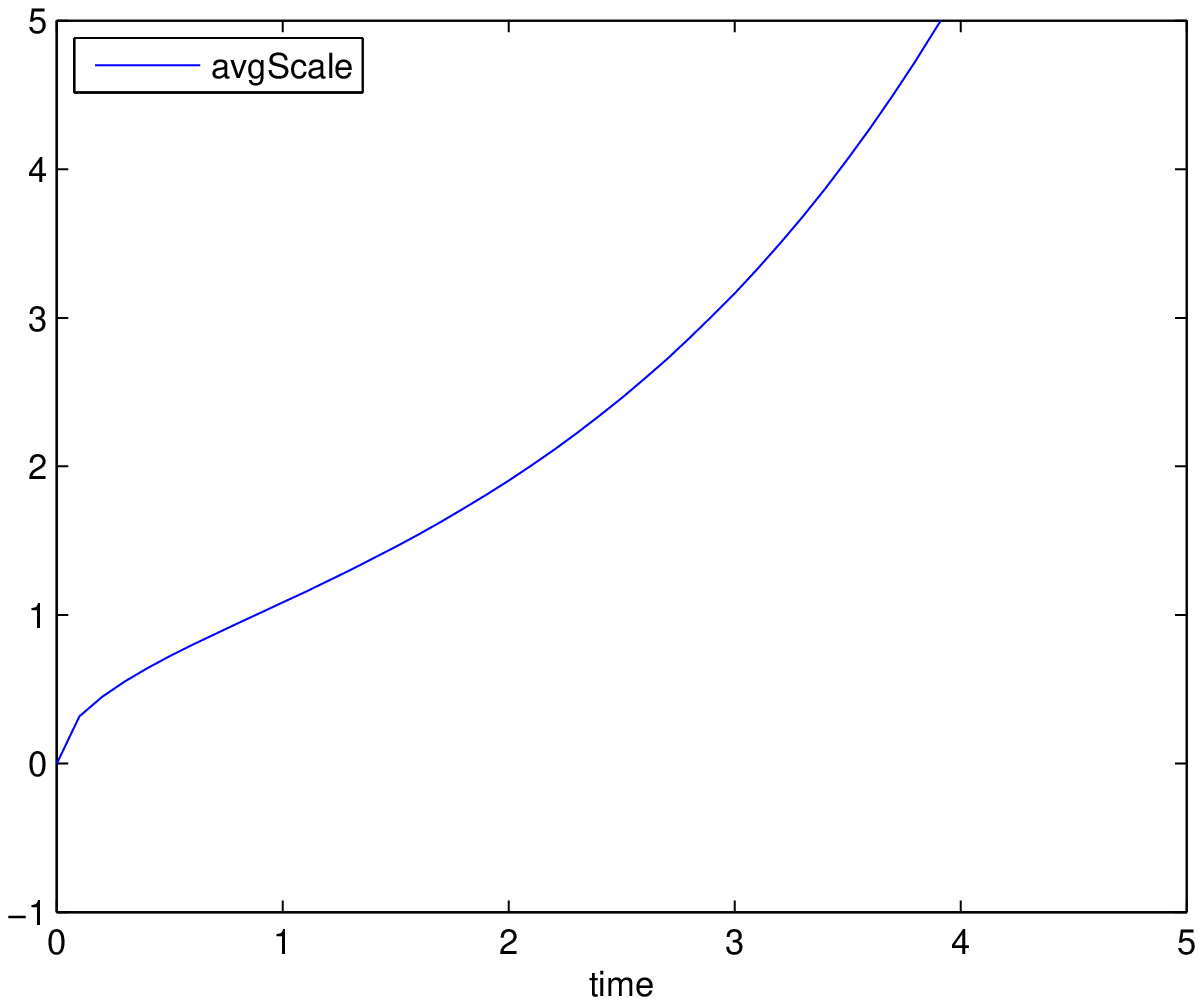}{0.45}{Variation of the average scale
factor $a$ for the parameter $n=2$ with time.}{0.45}
\vspace{1cm}
\noindent The time varying deceleration parameter can be obtained as

\begin{equation}
q=-\frac{a\ddot{a}}{\dot{a}^{2}}=n\left[1-(\tanh(\beta t))^{2}\right]-1.
\end{equation}
The time dependent deceleration parameter $(q)$ suggests the accelerating nature of the universe at present and decelerating in the past. This argument is recently supported by the observations of Type Ia supernova (Riess et al. [42,43]; Perlmutter et al.[44,45,46]; Tonry et al.[47]; Clocchiatti et al. [48]) and CMB anisotropies (Bennett et al.[49]; de Bernardis et al.[50]; Hanany et al.[51]). The value of the transition red-shift from decelerated expansion to accelerated expansion is about $0.5$. At present, the concept of constant deceleration is not acceptable as the deceleration parameter must show signature flipping [52-54]. The guage function $\phi$ [14, 15, 18] can be considered directly proportional to some constant power of average scale factor as,

\begin{eqnarray}
\phi=\phi_{0}a^{\alpha}=\phi_{0} sinh^{\alpha/n}(\beta t),
\end{eqnarray}
where $\alpha$=any constant and $\phi_{0}=$arbitrary constant.

\myfigures{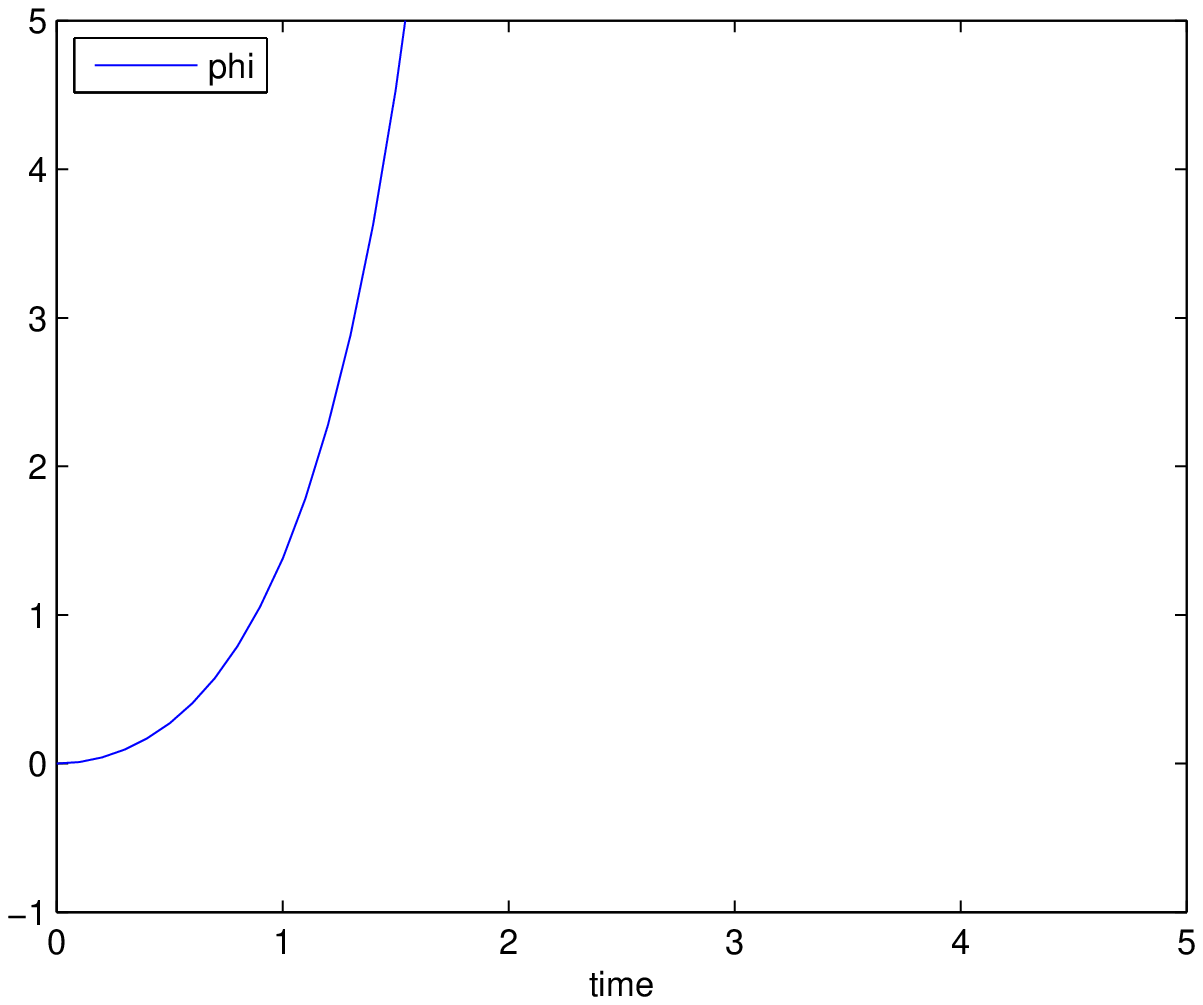}{0.45}{Variation of the gauge parameter $\phi$ for $n=0.5$
with time.} {0.45}{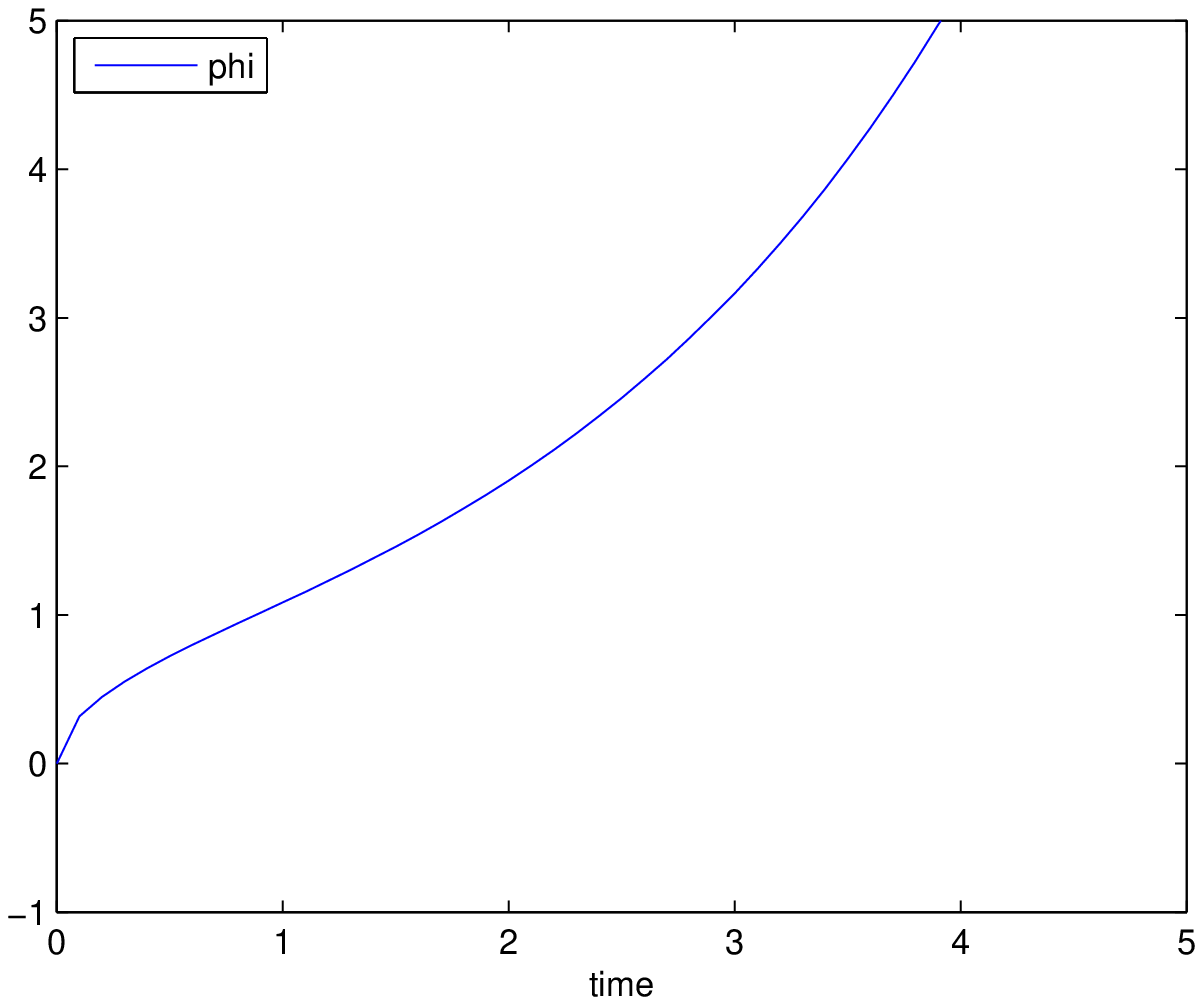}{0.45}{Variation of the gauge parameter $\phi$ for $n=2$
with time.}{0.45}

\section{Exact Solutions}
\noindent We solve the quadrature solution equations (20) to (22) by substituting in these equations the values of the average scale factor $a$ and guage function $\phi$ from equations (23) and (25) and integrating these equations to get the exact solutions for scale factors $A$, $B$ and $C$ as,
\begin{eqnarray}
A=\left[sinh (\beta t)\right]^{1/n},
\end{eqnarray}
\begin{eqnarray}
B= B_{0}\left[sinh (\beta t)\right]^{1/n} exp\left[\dfrac{sinh (2\beta t)-2\beta t}{4\beta}\right],
\end{eqnarray}
\begin{eqnarray}
C= C_{0} \left[sinh(\beta t)\right]^{1/n} exp\left[-\dfrac{sinh (2\beta t)-2\beta t}{4\beta}\right],
\end{eqnarray}
provided $\beta\neq0$. Here $M=1$ and $2\alpha+2n+3=0$. The expansion scaler $\theta$ is given by
\begin{eqnarray}
\theta = \dfrac{3\beta}{n}coth (\beta t).
\end{eqnarray}
\myfigures{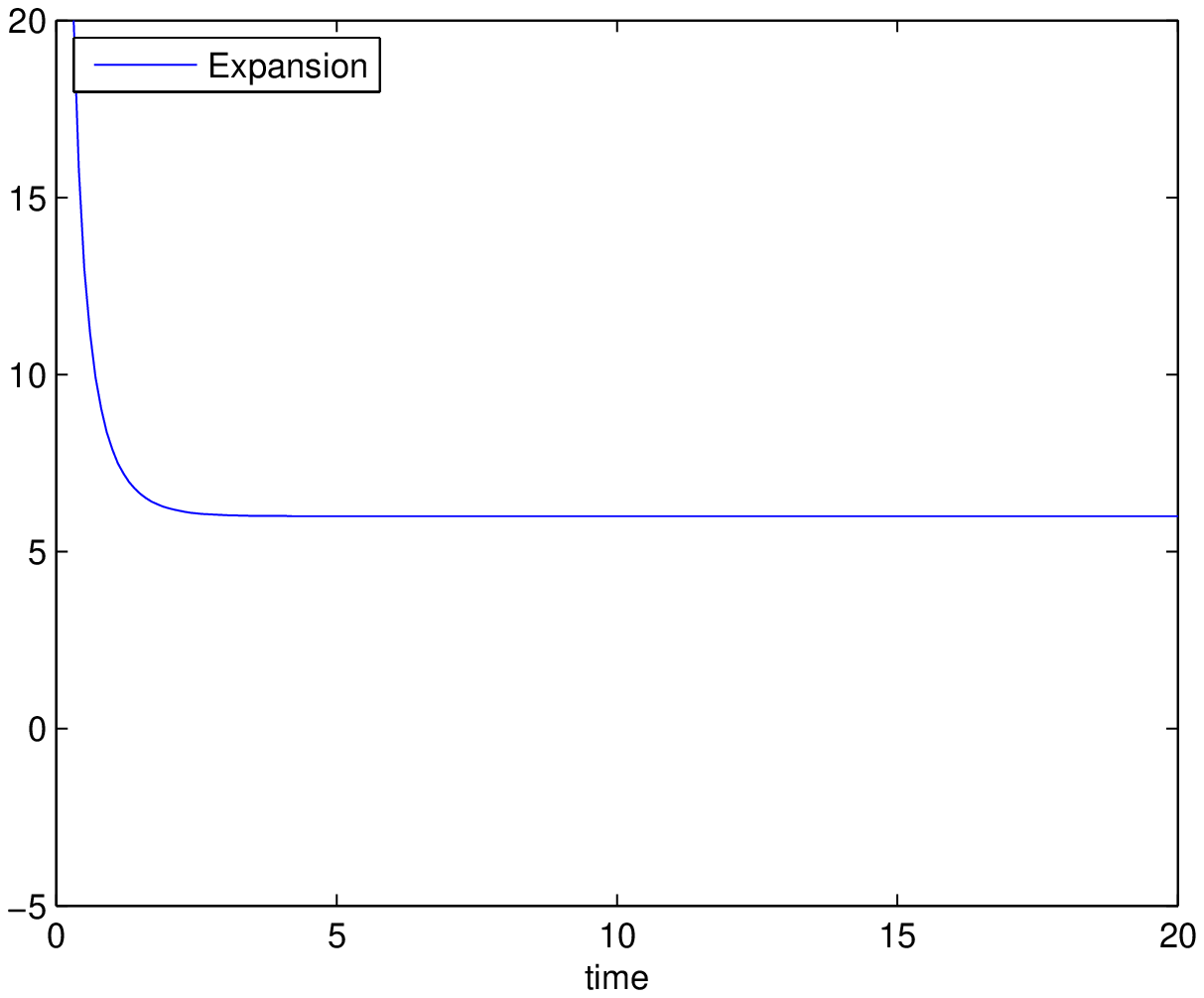}{0.45}{Behaviour of the Expansion scalar $\theta$
with time for $n=0.5$.} {0.45}{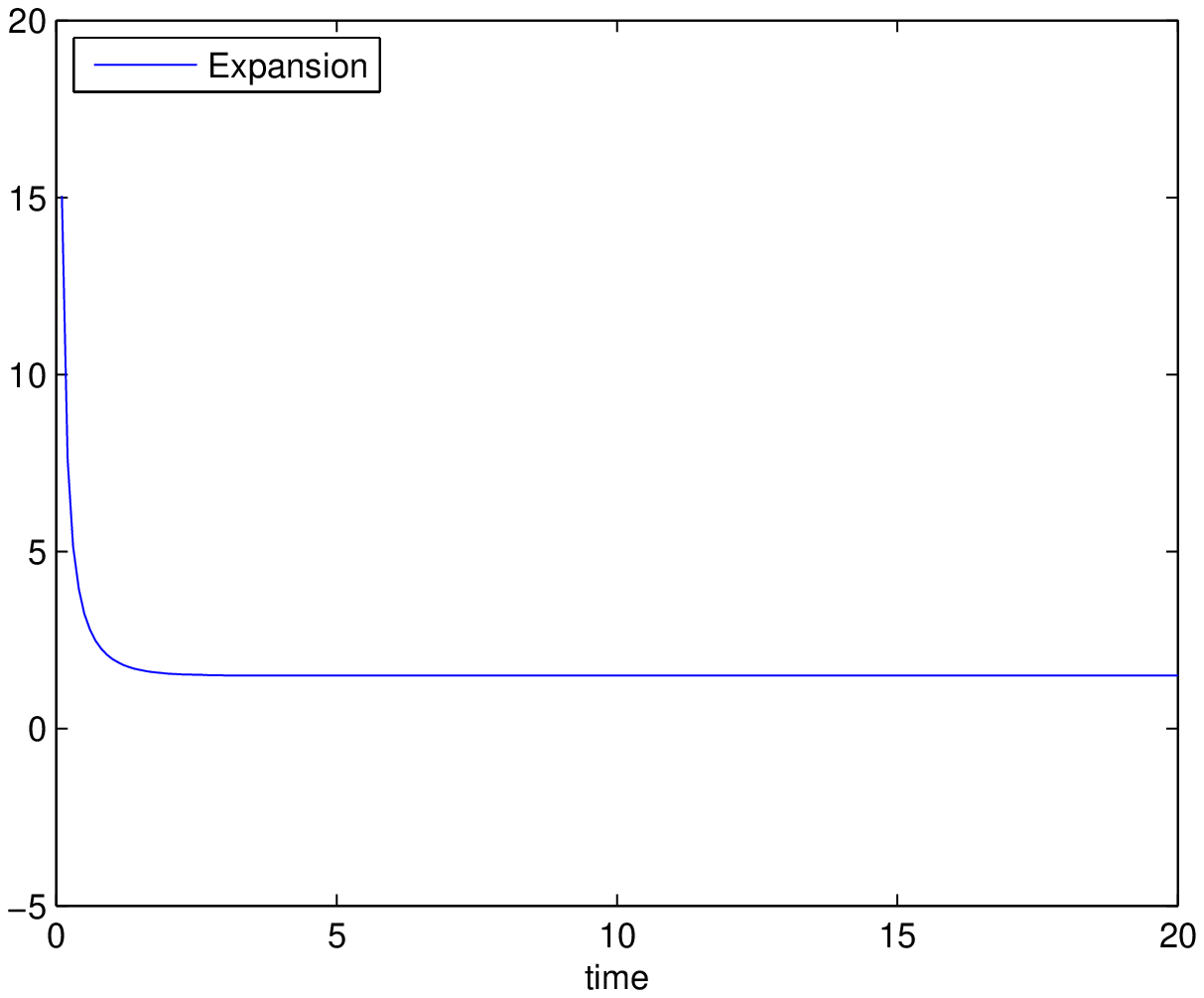}{0.45}{Behaviour of the Expansion scalar $\theta$ with time for $n=2$.}{0.45}

\noindent The shear scalar $\sigma^2$ can be obtained as

\begin{eqnarray}
\sigma^{2}=\dfrac{1}{8}[cosh (2\beta t)-1]^{2}-\dfrac{\beta ^{2}}{2n^{2}}coth^{2}(\beta t).
\end{eqnarray}

\myfigures{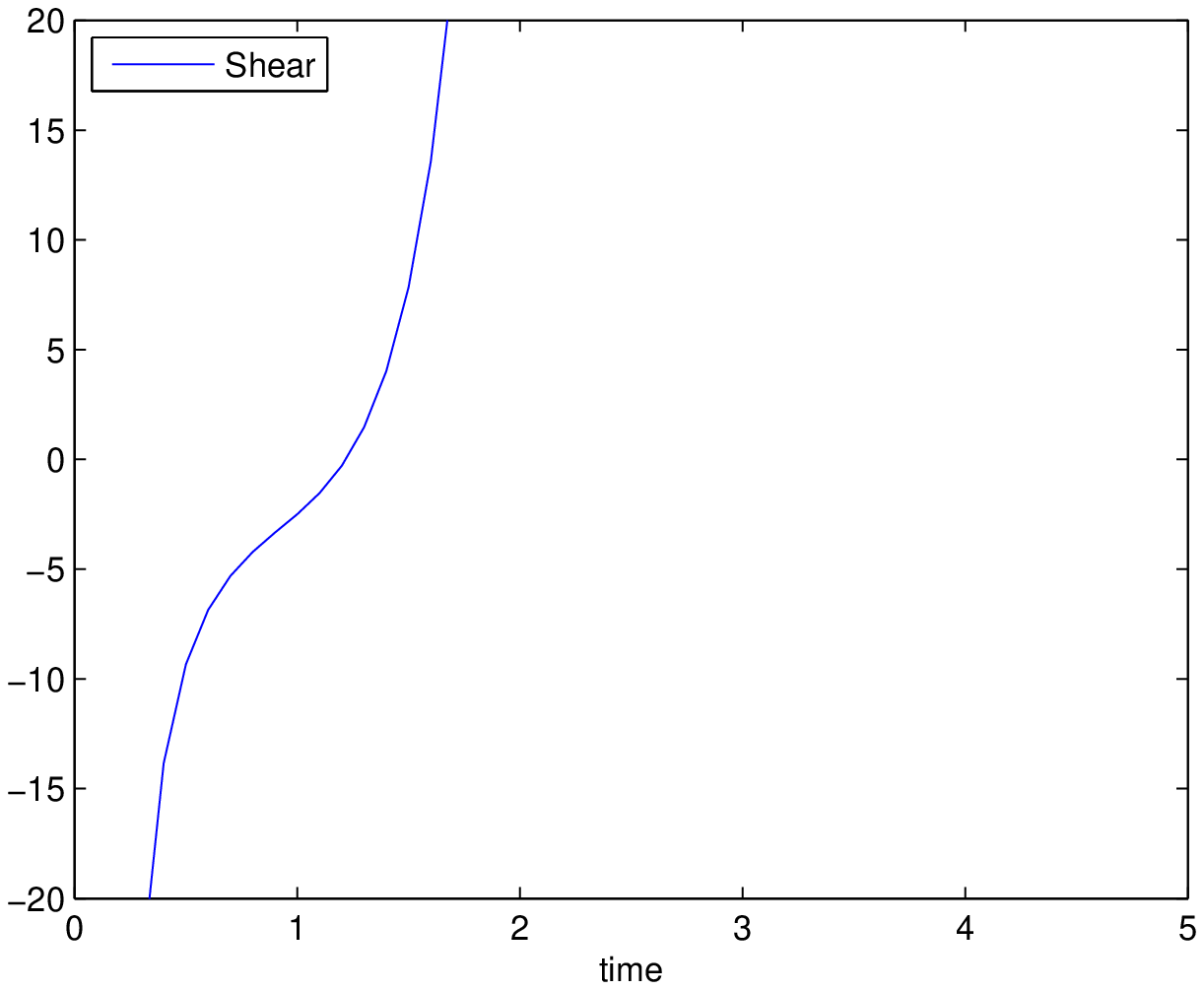}{0.45}{Variation of the shear scalar parameter $\sigma^{2}$ for $n=0.5$
with time.} {0.45}{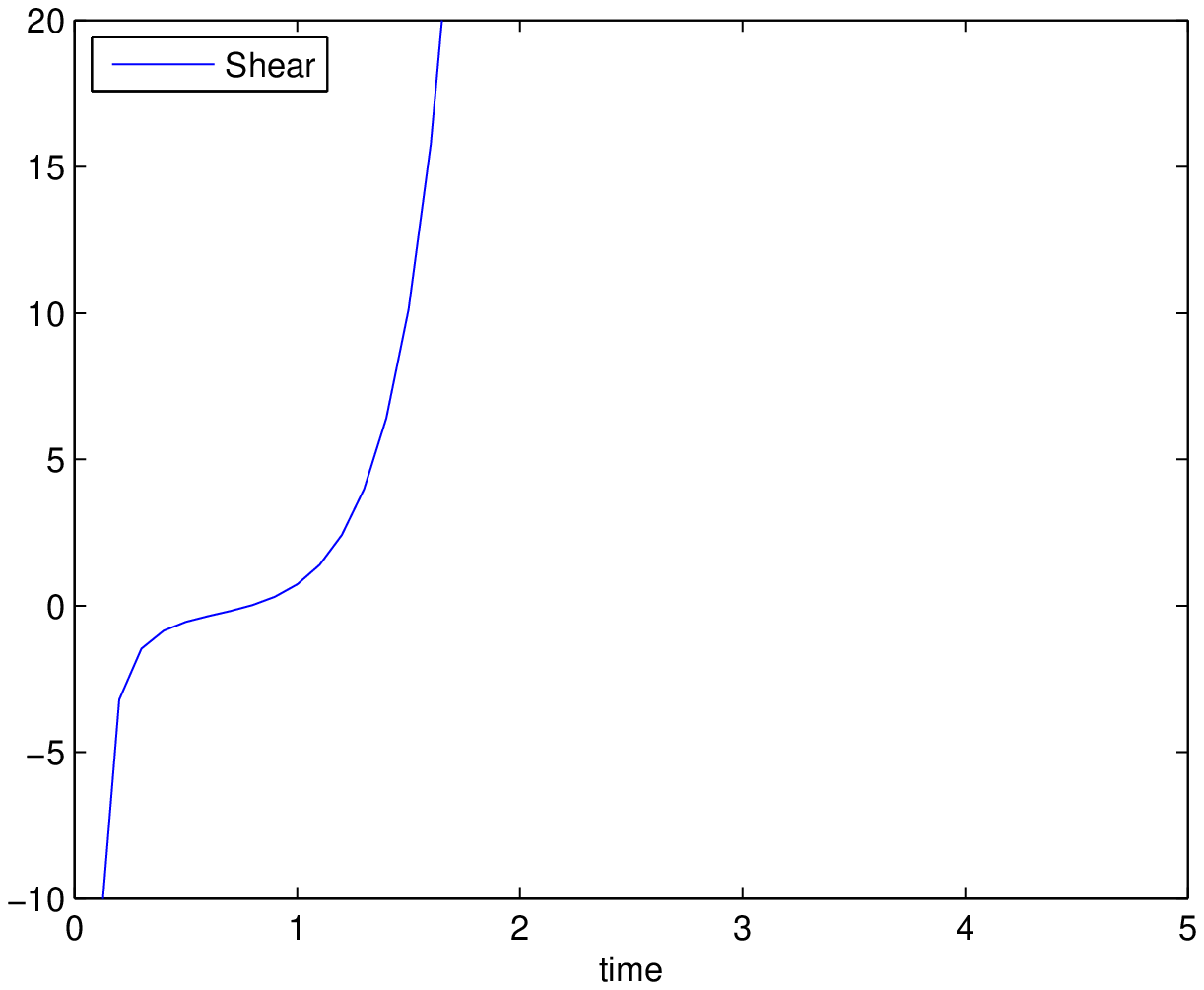}{0.45}{Variation of the shear scalar parameter $\sigma^{2}$ for $n=2$
with time.}{0.45}

\noindent The directional Hubble parameters $H_1$, $H_2$ and $H_3$ in the directions of $x-$, $y-$ and $z-$ coordinate axes respectively are calculated as follows
\begin{eqnarray}
H_1=\dfrac{\beta}{n}coth(\beta t),
\end{eqnarray}
\begin{eqnarray}
H_2=\dfrac{1}{2n}\left[2\beta coth (\beta t)+n(cosh(2\beta t)-1)\right],
\end{eqnarray}
\begin{eqnarray}
H_3=\dfrac{1}{2n}\left[2\beta coth (\beta t)-n(cosh (2\beta t)-1)\right].
\end{eqnarray}
The Hubble parameter $H$ can be obtained as
\begin{eqnarray}
H=\dfrac{\beta}{n}coth(\beta t).
\end{eqnarray}
\myfigures{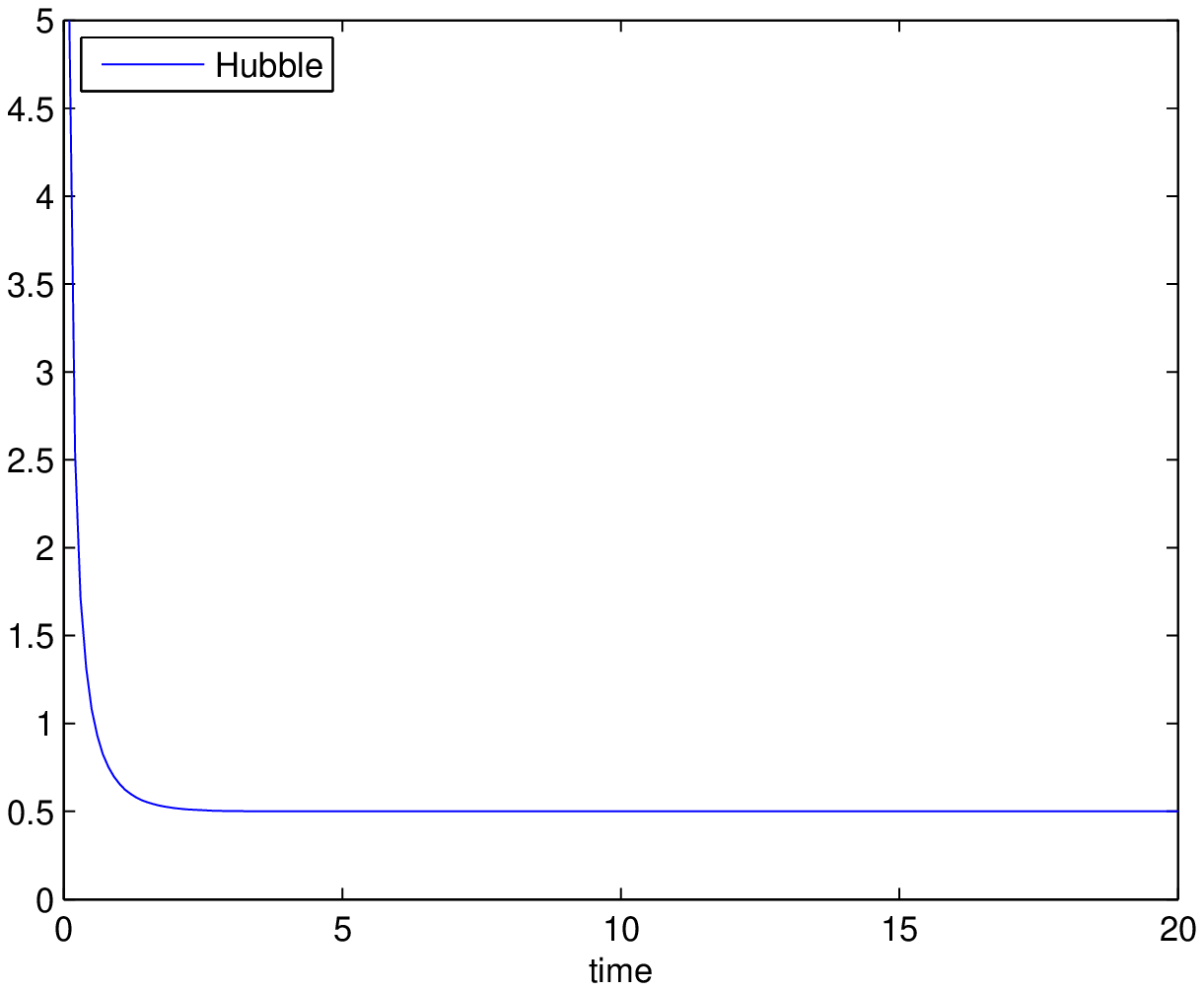}{0.45}{Variation of Hubble parameter $H$ for $n=0.5$
with time.} {0.45}{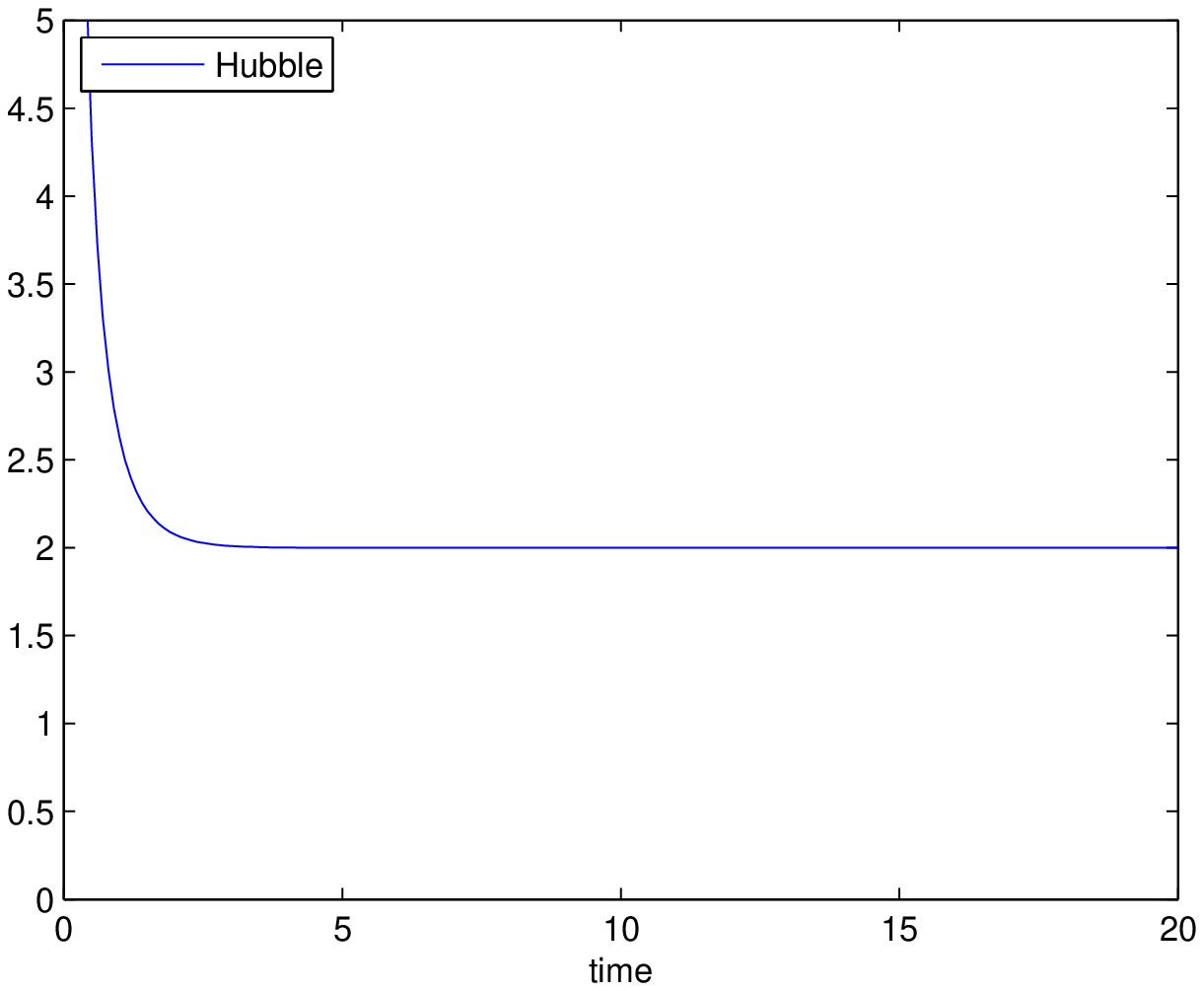}{0.45}{Variation of Hubble parameter $H$ for $n=2$
with time.}{0.45}

\noindent The volume scalar $V$ in this model is given by
\begin{eqnarray}
V=sinh^{3/n}(\beta t).
\end{eqnarray}

\myfigures{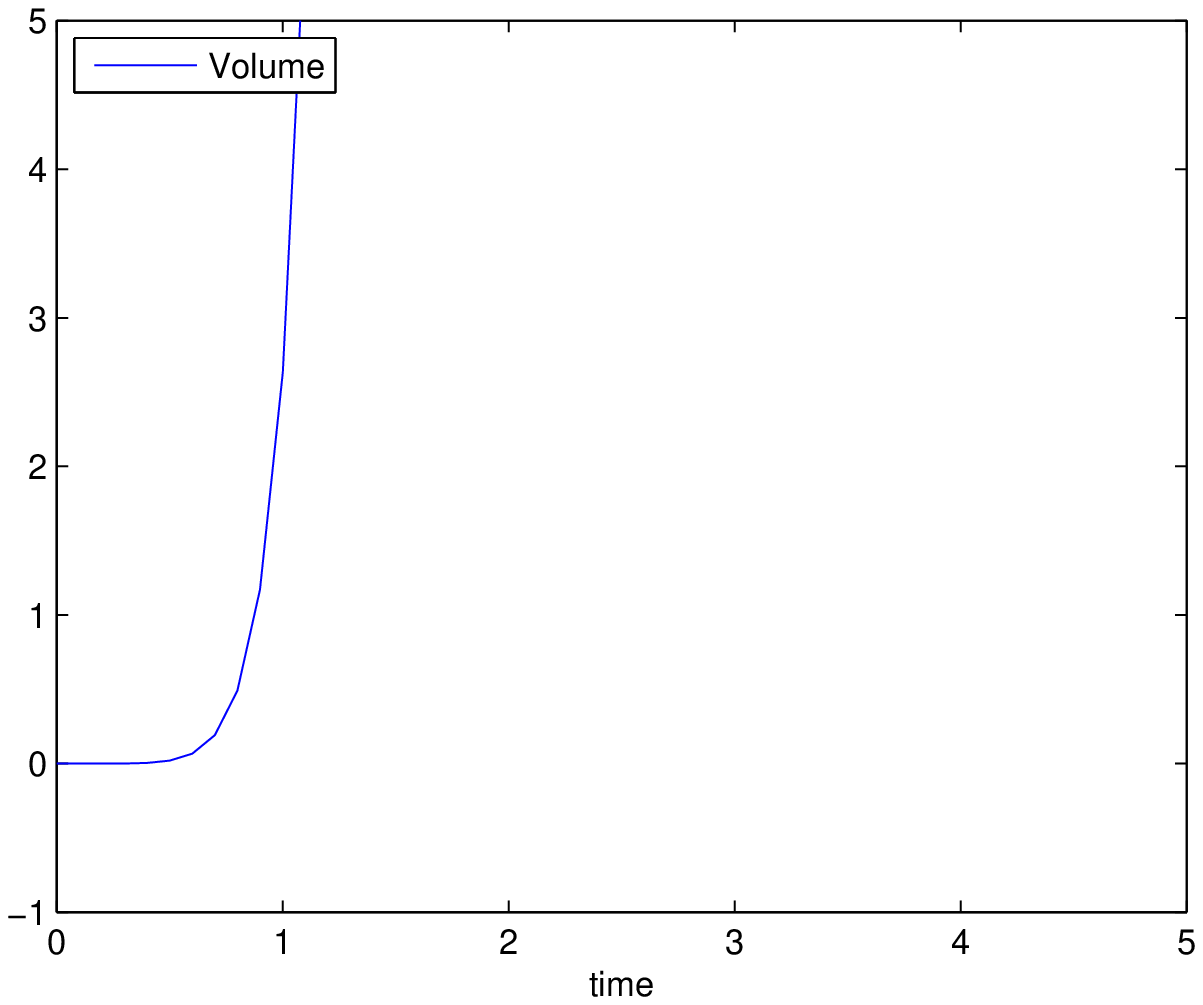}{0.45}{Behaviour of the Spatial Volume Parameter $V$ for $n=0.5$
with time.} {0.45}{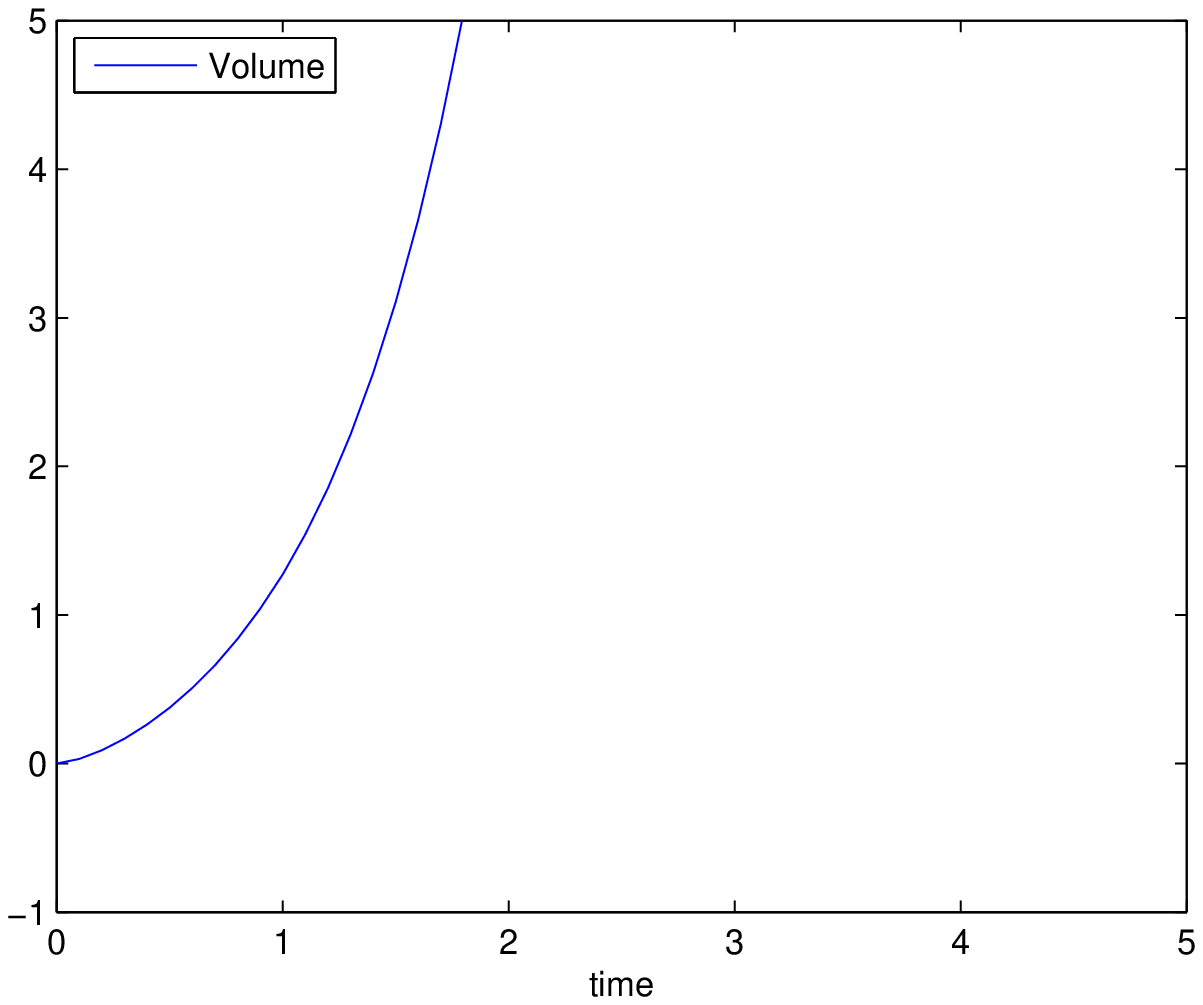}{0.45}{Behaviour of the Spatial Volume Parameter $V$ for $n=2$
with time.}{0.45}

\noindent The energy density $\rho$ and the pressure $p$ can be obtained as
\begin{eqnarray}
\rho = \dfrac{2\beta ^{2}}{n^{2}}(1+2\phi_{0}\alpha)coth ^{2}(\beta t)-\dfrac{1}{2}sinh ^{4}(\beta t)
  		-3 m^{2} [sinh^{-2/n}(\beta t)]-\phi_{0}\dfrac{\alpha \beta ^{2}}{n},
\end{eqnarray}
\begin{eqnarray}
p =\dfrac {\beta ^{2}}{n^{2}}\left[\phi_{0}\alpha(n-1)-\dfrac{5}{2}\right] coth^{2}(\beta t) +\dfrac {2\beta}{n}csch^{2}(\beta t)-\dfrac {1}{2}sinh ^{4}(\beta t)+m^{2} sinh^{-2/n}(\beta t)-\dfrac{\phi_{0}\alpha \beta ^{2}}{n}.
\end{eqnarray}

\noindent The behaviour of the above parameters can be thoroughly discussed. The average scale factor $a$ varies in such a manner that it starts with zero time $(t=0)$ and becomes infinity for the large time as $t\rightarrow \infty$. Also the accelerating and decelerating nature of the average scale can clearly be seen in Figures $1$ and $2$ for $0<n<1$ and $n>1$ respectively. Due to this value of average scale factor, the time varying deceleration parameter $q$ can be analyzed for its positive and negative values, i.e.,for $q>0$, $t<\frac{1}{\beta} \tanh^{-1}(1-\frac{1}{n})^{1/2}$ and $q<0$, $t>\frac{1}{\beta} \tanh^{-1}(1-\frac{1}{n})^{1/2}$. The volume scalar $V$ is zero at $t=0$ and it tends to infinity as $t\rightarrow \infty$. These behaviours of $V$ can be observed in figures $11$ and $12$. The accelerating and decelerating nature of the gauge function $\phi$ can be seen in figures $3$ and $4$. It has the same behaviour as that of average scale factor. The figures $5$ and $6$ show the variation of the expansion scalar. It will become infinity at $t=0$ but as $t\rightarrow \infty$, this parameter vanish. The behaviour of the shear scalar $\sigma^{2}$ can be observed from figures $7$ and $8$. The graphs in figures $9$ and $10$ give the variation of the Hubble parameter. The energy density $\rho$ and isotropic pressure $p$ will also diverge at $t=0$ and approach to zero for an infinite time. Here we conclude from the above results that the universe starts with zero volume and it has transition from initial anisotropy to isotropy at present epoch. So we can say that the model indicates a shearing, non rotating, expanding and accelerating universe with big bang singularity at its start. Here we observe that the model is accelerating for $0<n\leq 1$ and for $n>1$, the model evolves from a decelerating phase to an accelerating one. Recent observation SNe Ia supports the accelerating nature of the universe.

\section{Conclusion}
\noindent In this paper, we have obtained exact solutions for Scale Covariant theory of gravitation in Bianchi type V space-time. Here, we have used a time dependent deceleration parameter which yields the average scale factor $a(t)=Sinh^{1/n} (\beta t)$, where $n$ and $\beta$ are positive constants. The time dependent deceleration parameter supports the recent observation. The model represents an accelerating phase for $0<n\leq1$ and for $n>1$, there is a phase transition from early deceleration to a present accelerating phase. Following the concept of variable deceleration parameter, we get a singular cosmological model. Different physical and kinematical parameters have been obtained and thoroughly studied in all cases geometrically.

\vspace{0.5cm}

\noindent \textbf{References}

\vspace{0.5cm}

\noindent 1. G. Lyra, Math. Z. \textbf{54}, 52 (1951).

\vspace{0.1cm}

\noindent 2. W. D. Halford, Aust. J. Phys. \textbf{23}, 863 (1970).

\vspace{0.1cm}

\noindent 3. D. K. Sen, Z. Phys. \textbf{149}, 311 (1957).

\vspace{0.1cm}

\noindent 4. D. K. Sen and K. A. Dunn, J. Math. Phys. \textbf{12}, 578 (1971).

\vspace{0.1cm}

\noindent 5. A. Pradhan and H. R. Pandey, arXiv:gr-qc/0307038v1.

\vspace{0.1cm}

\noindent 6. A. Beesham, Aust. J. Phys. \textbf{41}, 833 (1988).

\vspace{0.1cm}

\noindent 7. T. Singh and G. P. Singh, J. Math. Phys. \textbf{32}, 2456 (1991).

\vspace{0.1cm}

\noindent 8. T. Singh and G. P. Singh, Nuovo Cimento B \textbf{106}, 617 (1991).

\vspace{0.1cm}

\noindent 9. T. Singh and G. P. Singh, Int. J. Theor. Phys. \textbf{31}, 1433 (1992).

\vspace{0.1cm}

\noindent 10. T. Singh and G. P. Singh, Fortschr. Phys. \textbf{41}, 737 (1993).

\vspace{0.1cm}

\noindent 11. Shri Ram and P. Singh, Int. J. Theor. Phys. \textbf{31}, 2095 (1992).

\vspace{0.1cm}

\noindent 12. G. P. Singh and K. Desikan, Pramana J. Phys. \textbf{49}, 205 (1997).

\vspace{0.1cm}

\noindent 13. A. Pradhan et al., Int. J. Mod. Phys. D \textbf{10}, 339 (2001).

\vspace{0.1cm}

\noindent 14. V. Canuto et al., Phys. Rev. Lett. \textbf{39}, 429 (1977).

\vspace{0.1cm}

\noindent 15. V. Canuto et al., Phys. Rev. D \textbf{16}, 1643 (1977).

\vspace{0.1cm}

\noindent 16. C. M. Wesson., Gravity Particles and Astrophysics(New York. Reidel)\\
\indent Dordrecht Holland(1980).

\vspace{0.1cm}

\noindent 17. C. M. Will., Phys. Rep. \textbf{113}, 345 (1984).

\vspace{0.1cm}

\noindent 18. Shri Ram et al.,  Chin. Phys. Lett. \textbf{26}, No.8, 089802(2009).

\vspace{0.1cm}

\noindent 19. M. Zeyauddin and B. Saha.,  Astrophys. Space. Sci. \textbf{343}, 445 (2013).

\vspace{0.1cm}

\noindent 20. D. R. K. Reddy et al.,  Astrophys. Space. Sci. \textbf{307}, 365 (2007).

\vspace{0.1cm}

\noindent 21. D. R. K. Reddy et al.,  Astrophys. Space. Sci. \textbf{204}, 155 (1993).

\vspace{0.1cm}

\noindent 22. A. Beesham.,  J. Math. Phys.  \textbf{27}, 2995 (1986).

\vspace{0.1cm}

\noindent 23. R. Venkateswarlu and P. K. Kumar.,  Astrophys. Space. Sci. \textbf{298}, 403 (2005).

\vspace{0.1cm}

\noindent 24. B. Saha., V. Rikhvitsky., Physica. D \textbf{219}, 168 (2006).

\vspace{0.1cm}

\noindent 25. M. Zeyauddin., B. Saha., Astrophys. Space. Sci. \textbf{343}, 445 (2013).

\vspace{0.1cm}

\noindent 26. S.Ram, M. Zeyauddin, C.P.Singh, Int. J. Mod. Phys. A \textbf{23}, 4991(2008).

\vspace{0.1cm}

\noindent 27. S.Ram, M.K.Verma, M. Zeyauddin,  Chin. Phys. Lett. \textbf{26}, 089802(2009).

\vspace{0.1cm}

\noindent 28. S. Ram, M. Zeyauddin, C.P. Singh, J. Geom. Phys.  \textbf{60}, 1671(2010).

\vspace{0.1cm}

\noindent 29. C.P. Singh, S.Ram, M. Zeyauddin,  Astrophys. Space. Sci. \textbf{315}, 181(2008).

\vspace{0.1cm}

\noindent 30. A. Pradhan, Indian. J. Phys \textbf{88(2)}, 215(2014).

\vspace{0.1cm}

\noindent 31. N. Ahmed, A. Pradhan,  Int.J.Theor.Phys \textbf{53}, 289(2014).

\vspace{0.1cm}

\noindent 32. A. Pradhan, S. Otarod,  Astrophys. Space Sci \textbf{306}, 11(2006).

\vspace{0.1cm}

\noindent 33. O. Akarsu, T. Dereli, Int. J. Theor. Phys. \textbf{51}, 612 (2012).

\vspace{0.1cm}

\noindent 34. A. Pradhan, R. Jaiswal, K. Jotania, R. K.  Khare, Astrophys. Space Sci \textbf{337}, 401 (2012).

\vspace{0.1cm}

\noindent 35. C, Chawla, R. K.  Mishra, Rom. J. Phys \textbf{58}, 75 (2013).

\vspace{0.1cm}

\noindent 36. C. Chawla, R. K. Mishra, A. Pradhan, Eur. Phys. J. Plus \textbf{127}, 137 (2012).

\vspace{0.1cm}

\noindent 37. R. K. Mishra, A. Pradhan, C. Chawla, Int. J. Theor. Phys \textbf{52}, 2546 (2013).

\vspace{0.1cm}

\noindent 38. A. Pradhan, Res. Astron. Astrophys \textbf{13}, 139 (2013).

\vspace{0.1cm}

\noindent 39. A. Pradhan, A. K. Singh, D. S. Chouhan,  Int. J. Theor. Phys \textbf{52}, 266 (2013).

\vspace{0.1cm}

\noindent 40. A. Pradhan, R. Jaiswal, R. K. Khare, Astrophys. Space Sci \textbf{343}, 489 (2013).

\vspace{0.1cm}

\noindent 41. H. Amirhashchi, A. Pradhan, H. Zainuddin, Res. Astron. Astrophys \textbf{13}, 129 (2013).

\vspace{0.1cm}

\noindent 42. A.G. Riess et al.,  Astron. J. \textbf{116}, 1009 (1998).

\vspace{0.1cm}

\noindent 43. A.G. Riess et al.,  Astrophys. J. \textbf{607}, 665(2004).

\vspace{0.1cm}

\noindent 44. S. Perlmutter et al.,  Nature \textbf{391}, 51 (1998).
\vspace{0.1cm}

\noindent 45. S. Perlmutter et al.,  Astrophys. J., \textbf{517}, 565 (1999).

\vspace{0.1cm}

\noindent 46. S. Perlmutter et al.,  Astrophys. J., \textbf{598}, 102 (2003).

\vspace{0.1cm}

\noindent 47. J. L. Tonry et al., Astrophys. J. \textbf{594}, 1 (2003)

\vspace{0.1cm}

\noindent 48. A. Clocchiatti et al., Astrophys. J. \textbf{642}, 1 (2006).

\vspace{0.1cm}

\noindent 49. C.L. Bennett et al.,  Astrophys. J. Suppl. \textbf{148}, 1 (2003).

\vspace{0.1cm}

\noindent 50. P. de Bernardis et al., Nature \textbf{666}, 716 (2007).

\vspace{0.1cm}

\noindent 51. S. Hanany et al., Astrophys. J. \textbf{545}, L5 (2000).

\vspace{0.1cm}

\noindent 52. T. Padmanabhan, T. Roychowdhury, Mon. Not. R. Astron. Soc. \textbf{344}, 823 (2003).

\vspace{0.1cm}

\noindent 53. L. Amendola, Mon. Not. R. Astron. Soc. \textbf{342}, 221 (2003).

\vspace{0.1cm}

\noindent 54. A.G. Riess et al., Astrophys. J. \textbf{560}, 49 (2001).

\vspace{0.1cm}

\end{document}